\documentclass[journal]{IEEEtran}
\usepackage[utf8]{inputenc}
\usepackage[english]{babel}
\usepackage{graphicx}
\usepackage{amsthm}

\usepackage{setspace}
\usepackage{bibentry} 
\usepackage{xcolor,soul,framed} 
\usepackage{cite}
\usepackage{upgreek}
\colorlet{shadecolor}{yellow}
\graphicspath{{../pdf/}{../jpeg/}}
\DeclareGraphicsExtensions{.pdf,.jpeg,.png,.bmp}
\usepackage{amsmath,amssymb}
\usepackage{mathabx}
\usepackage{array}
\usepackage{mdwmath}
\usepackage{mdwtab}
\usepackage{eqparbox}
\usepackage{url}
\usepackage{amsmath}
\usepackage{breqn}
\usepackage[colorlinks]{hyperref}

\begin{document}
\bstctlcite{IEEEexample:BSTcontrol}
   \title{UAVs-Enabled Maritime Communications: Opportunities and Challenges}
   \author{Muhammad Waseem~Akhtar,~\IEEEmembership{Graduate Student Member,~IEEE,}  Nasir~Saeed,~\IEEEmembership{Senior Member,~IEEE,}
   \thanks{M. W. Akhtar is with the  School of Electrical Engineering and Computer  Science (SEECS), National University of Sciences and Technology(NUST), Islamabad, Pakistan (e-mail: makhtar.dphd17seecs@seecs.edu.pk).
   N. Saeed is with the Department
of Electrical Engineering, King Abdullah University of Science and Technology, Thuwal, Makkah, Saudi Arabia (e-mail: mr.nasir.saeed@ieee.org)}}
 \maketitle
\begin{abstract} 
The next generation of wireless communication systems will integrate terrestrial and non-terrestrial networks targeting to cover the undercovered regions, especially connecting the marine activities. Unmanned aerial vehicles (UAVs) based connectivity solutions offer significant advances to support the conventional terrestrial networks. However, the use of UAVs for maritime communication is still an unexplored area of research. Therefore, this paper highlights different aspects of UAV-based maritime communication, including the basic architecture, various channel characteristics, and use cases. The article, afterward, discusses several open research problems such as mobility management, trajectory optimization, interference management, and beamforming. 
\end{abstract} 


\IEEEpeerreviewmaketitle


\section{Introduction}
The seawater covers around 70$\%$ of planet Earth where over 90$\%$ of the world’s products are moved by a commercial fleet of approximately 46,000 ships \cite{uav1}. The world is experiencing an ever-growing booming marine economy with continuous development in conventional sectors such as fisheries and transportation and exploring dimensions in maritime activities such as tourism, exploring oil and gas resources, and weather monitoring. Most of these applications depend on a reliable and efficient maritime communication network. \par 
The existing maritime networks comprise mainly either too low bandwidth, very high frequency (VHF) radios, or too high-cost satellite communication networks to support the international maritime safety organization (IMO) eNavigation concept. However, the emerging maritime networks need wideband low-cost communication systems to achieve better security, surveillance, and coverage for efficient working conditions for the onboard crew and passengers. Although wireless broadband access (WBA) can fulfill the IMO eNavigation requirement, the implementation of WBA technologies in maritime areas is questionable.\par
The typical marine networks comprise a mesh network of different entities in an integrated satellite-air-sea-ground network. A stand-alone satellite-based solution considerably boosts its potential to cover a large area with high-speed data transmission. However, it suffers from unavoidable large propagation delays and expensive implementation costs. Alternatively, HF/VHF-based systems are simple to implement but have limited utilization, i.e., only in vessel identification, tracking/monitoring, and alerting. Therefore, it is of great importance to develop high-speed maritime networks to improve the onboard user experience. As a result, maritime communications have garnered substantial interest in the recent past, where the primary purpose is to enhance the broadband network coverage for the terrestrial users with the aid of unmanned aerial vehicles (UAVs), that can serve as aerial base stations and relays \cite{uav2,akhtar2020shift}. \par
\begin{figure}[t]
\centerline{\includegraphics[width=0.5\textwidth]{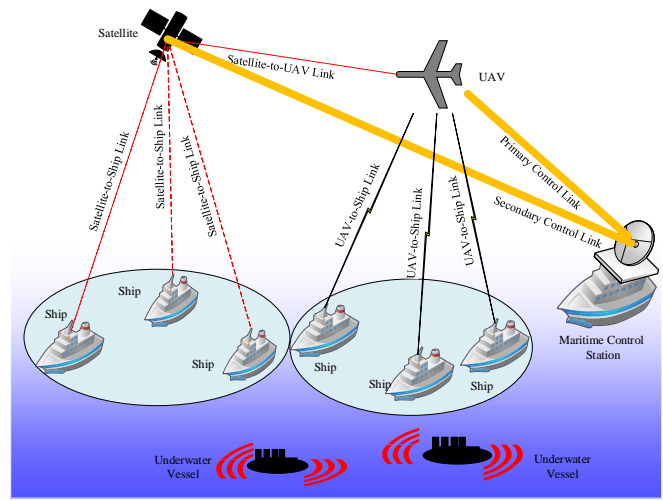}}
 \caption{A depiction of basic network architecture for UAV-aided maritime communication}\label{sea2}
\end{figure}
In this context, UAVs can play a vital role in maritime communications either as relays or flying sensors gathering information in cheaper, safer, and faster ways. For example, the UK Royal Navy actively uses UAVs to identify the defects in ships to diagnose and resolve the issues while keeping ships in the sea, reducing maintenance costs and time. Moreover, UAVs can also be helpful for maritime natural resource exploration purposes such as oil and gas exploration, especially in harsh and challenging environmental conditions. Furthermore, UAVs equipped with high-resolution cameras can also be used for security and surveillance purposes. A single drone can gather more information than the camera installed at different locations. Inspired by these trends, we present the key aspects of UAVs-aided maritime communication networks. Our major contributions in this paper are summarized as follows:
\begin{itemize}
\item	First, we present a design architecture of UAV-based maritime communication network
\item	Then, we discuss the channel characteristics in maritime communication networks, such as air-to-sea, and near sea surface channels. Also, we present the use-cases of UAV-aided maritime communication
\item	Finally, we present the research challenges and future directions for UAV-based maritime communication networks.   
\end{itemize}
\section{UAV-Aided Maritime Communication Network Architecture}
The basic network architecture of a UAV-aided maritime communication network is shown in Fig.~\ref{sea2}. UAVs are simultaneously connected with the maritime control station (MCS), satellite, and sea vessels in such a network. The communication links between UAVs and MCS, satellites, and ships are primary, whereas the communication link between satellite and MCS is secondary. In the following, we discuss the MCS, control links, and data links in detail.
\subsection{Maritime Control Station}
 A marine control station is the brain of maritime networks positioned on the ship, UAVs, or underwater to facilitate the operators of UAVs.  The control station may be either stationary or movable for command and control (C\& C) transmission. The control station equipment can be as simple as a laptop and antenna installed to it and as complex as a rat's nest with wires, antenna, computers, electronics boxes, joysticks, and monitors. 
\subsection{Control Links}
The link from a base station to users, in UAVs, assisted maritime networks, is called the control link. The control link is responsible for the transmission of commands and controls from a base station to the users in the uplink.
\subsection{Data Links}
Information is exchanged in the maritime networks using data links where the communication technologies are responsible for data delivery between system elements and external units. The fundamental challenges of the maritime network are the security of C\& C from a base station to the users, cognitive control of bandwidth, frequency, and data flow. Following are the different types of data links that exist in maritime networks.
\subsubsection{UAV-Ship and Satellite-Ship Data Links}
These links deliver information from the UAV/Satellite to a sea-based reception device. These links are responsible for the data communication between UAVs-and-ships, and satellite-and-ships.
\subsubsection{UAV-Satellite, UAV-UAV, and Satellite-Satellite Data Links}
UAVs can cooperate with other space/airborne platforms, such as satellites and other UAVs. This types of data links demand to establish air-to-air communication between the platforms. The establishment of these links is more challenging due to the relative movement of both transmitters and receivers \cite{saeed2021point}.
\begin{figure}[t]
\centerline{\includegraphics[width=0.5\textwidth]{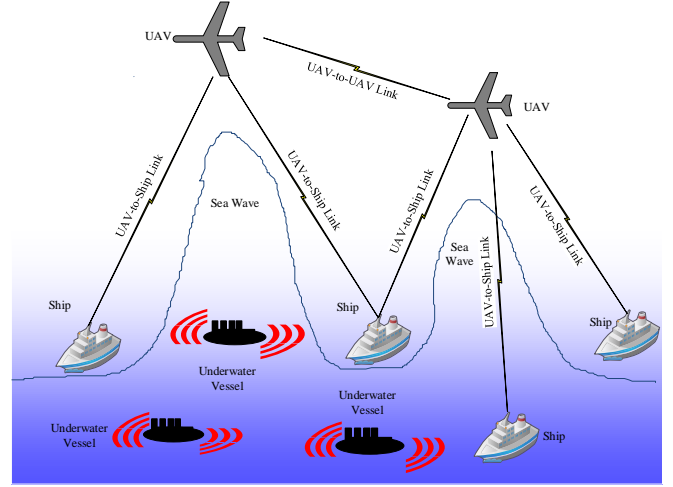}}
 \caption{UAV as a use case of reliable maritime communication in dynamically changing environmental conditions.}\label{sea1}
\end{figure}
\section{Channel Characteristics}
To establish an efficient maritime communication network mentioned above, it is important to comprehend and model the wireless channels. As far as maritime communication is concerned, there are three major channel types to be investigated. First is an air-to-sea channel, that is used for the communication between UAV and ships. The second is a near-sea-surface channel, that is used for ship-to-ship communication. And finally, an underwater communication channel, that is used for the communication between the underwater vessels. Underwater communication channels can further be divided into near-sea-surface (i.e., up to 600m below the sea surface) and deep-sea underwater (i.e., more than 600m below the sea surface) wireless channels due to differences in their characteristics, such as temperature, salinity, and atmospheric pressure at different sea level.\par
Maritime wireless channel differs from the conventional terrestrial channels by many aspects, such as ducting effect and heavy scattering over the sea surface, unpredictable sea wave proportions, water density, and temperature variations in the sea. All of these aspects result in significant complexity in the receiver design. Although the satellite-to-ship channels have been explored extensively in the past \cite{uav13}, the wireless channels in the terrestrial and non-terrestrial integrated networks (TaNTIN) \cite{akhtar2021tantin} are less explored for the near coast situation. Therefore, recently researchers have started investigating maritime wireless channels and developed several models.  \par
The maritime wireless channels' two most essential and distinguishing properties are sparsity and location dependence. The sparsity is extensively observed in the maritime environment especially for the unpredictable scattering and maritime receivers' distribution, whereas the location dependency feature applies that there should be a completely different channel model for different locations of the maritime receiver. Fig.~\ref{sea1} depicts the channel variations observed at the sea level due to the travelling sea waves, moving UAVs, and ships. Sea waves traveling in random directions, and with dynamic wave amplitudes cause a high level of fluctuations in signal-to-interference-plus-noise ratio (SINR) level at the receiver. Similarly, the variable speed of UAVs and ships leads to an unpredictable Doppler effect. Consequently, due to these traits, new difficulties and opportunities develop in the maritime communication system design.  \par
In the following, we discuss different models for the air-to-sea, near-sea-surface, and underwater wireless channels.
\begin{figure}[t]
\centerline{\includegraphics[width=0.5\textwidth]{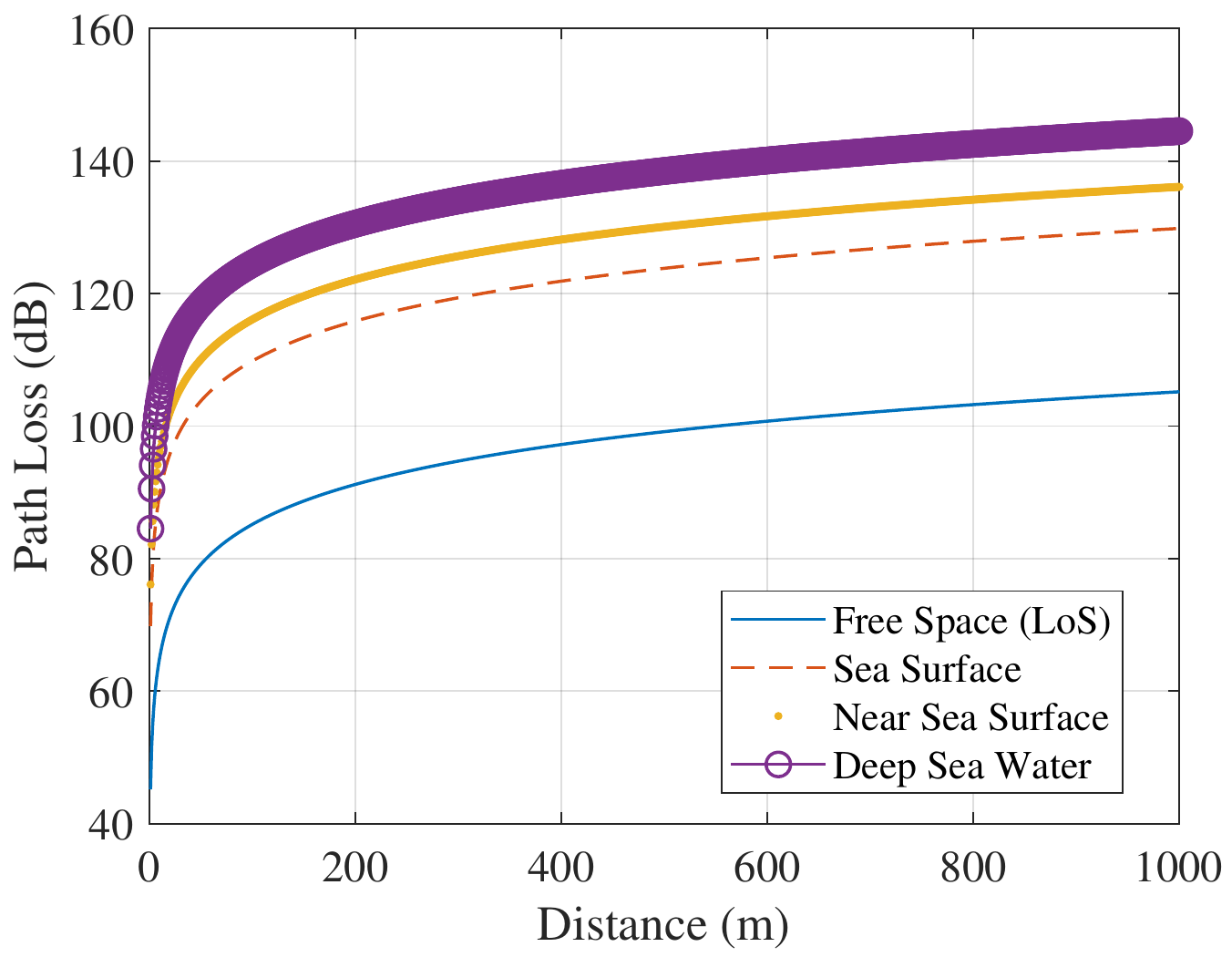}}
 \caption{A depiction of path-loss for free space (LoS), sea-surface, near sea-surface, and deep sea water channels at 500 kHz acoustic waves.}\label{AC_500}
\end{figure}
\subsection{Air-to-Sea Channel}
The air-to-ground channels are widely studied in the literature \cite{uav14}. However, the air-to-sea channel differs from the air-to-ground channel in many aspects due to the differences such as ducting and sparsity effect, and instability in the maritime environment, which leads to the remarkable difference in channel modeling. Usually, in many cases, the two-ray model is applied. The first component of the two-rays model is the line-of-sight (LoS) component, and the second is the surface-reflected ray component. When the transmission distance is very large and the transmitter is located at some notable height, the curve-Earth two rays (CE2R) model is used to cater for the Earth curvature \cite{uav15}. \par
In some cases, the rays received from other weak scattered paths can also be considered, apart from two strong paths. However, a dispersion around the maritime receiver is observed when the transmitter is located at a very high altitude \cite{uav16}. As compared to the terrestrial (i.e., near the urban area) environment, a maritime receiver is expected to face more sparse scattering, which may lead to simplification in the air-to-sea channel modeling. \par
As discussed earlier, a standard two rays or three rays model can be used in an air-to-sea channel. However, due to long-distance transmission in the maritime environment, two main elements i.e., ducting effect and the Earth curvature must be taken into account. Subsequently, this necessitates the application of the CE2R model in the maritime environment, which has a different path-loss model as compared to the flat-Earth assumption for short-distance transmission in the terrestrial communication systems.  Also, the location of the transmitter (UAV or satellite) is usually above the ducting layer, therefore, a part of the radio energy could be absorbed in the ducting layer especially when the gazing angle (the angle between the sea surface and the direct path) is less than a threshold. In this case, the ray trapping action of the ducting layer can also increase the power of the received signal, resulting in reduced path loss \cite{wang2018wireless}.
\begin{figure}[t]
\centerline{\includegraphics[width=0.5\textwidth]{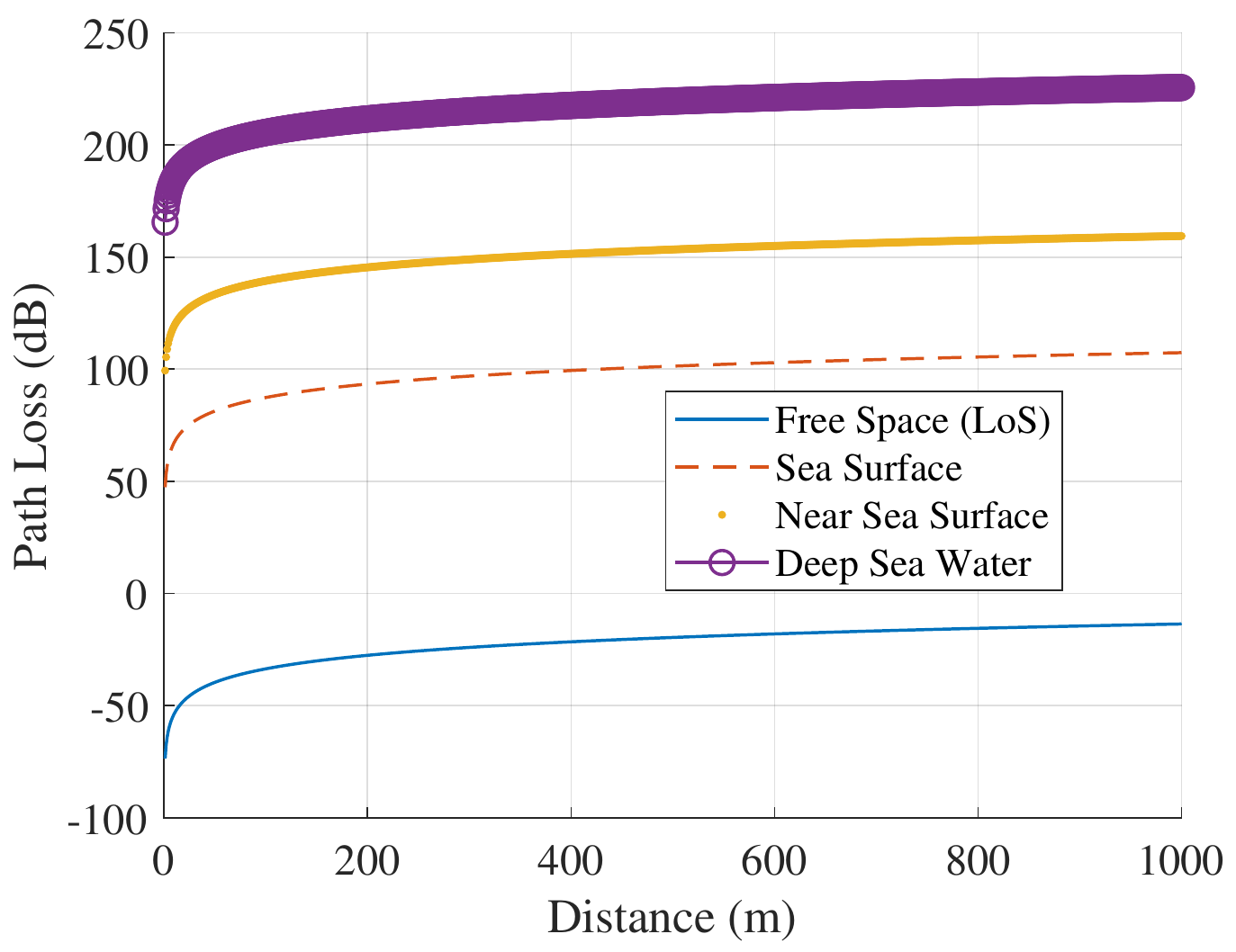}}
 \caption{A depiction of path-loss for free space (LoS), sea-surface, near sea-surface, and deep sea water channels at 500 kHz radio frequency  waves.}\label{RF_500}
\end{figure}
\subsection{Near-Sea-Surface Channel}
As motioned earlier, near-sea-surface (such as ship-to-ship, ship-to-land, and land-to-ship) channels are distance-dependent. Different channel models can be used for different locations of transmitter and receivers. The standard two-ray model can be used for a modest distance between transmitter and receiver. However, the LoS component and the reflected ray component vanish away due to Earth curvature with the increase in the distance between transmitter and receiver. However, the receiver can still receive the signal transmitted due to the ducting effect, provided the proper beam alignment between transmitter and receiver. Conclusively, as the distance between transmitter and receiver increases, the two-ray or three-ray channel model is replaced by duct-only in the end. The ducting effect across the sea surface allows Beyond LoS (BLoS) transmission in marine communications, which has gained much popularity in secure and long-distance maritime communication. \\
Fig.~\ref{AC_500} shows the path-loss \cite{wang2019seawater,morozs2020channel,wang2018wireless} against the distance between transmitter and receiver for different maritime channels with acoustic waves at 500 kHz frequency. It is clear that the path-loss varies with the level of water density in the wireless channel. For instance, the path-loss at the deep seawater is higher than free space, near the sea surface, and sea surface  channels. The reasons for this are the factors of temperature, shadowing, and density of the water. We also show the trend of path loss for radio-frequency waves in Fig.~\ref{RF_500}, where the radio frequencies face the highest path loss in deep-sea water channels as compared to other maritime wireless channels. By comparing Fig.~\ref{AC_500} and Fig.~\ref{RF_500}, we can analyze that acoustic waves are more suitable for maritime under communication in the underwater environment. At the same time, RF is better suited for near-surface and free-space links.
\subsection{Underwater Channel}
Besides air-to-sea and near sea-surface wireless channels, underwater wireless channel modeling is also exciting to replace the aquatic cabled systems. For example, remotely controlled vehicles are connected to the mother ship by a cable that can run over several kilometers and give tremendous power to the remote end, along with high-speed communication messages. However, to improve the flexibility and mobility of underwater networks, modern remotely controlled vehicles can communicate wirelessly. Today, underwater communications primarily use acoustic technology with other complementary technologies, such as RF, optical communication, or even magnetic induction for short-range networks (usually 1–10m) \cite{Khalil2021}. Several existing works discuss channel modeling and networking for underwater networks; however, embedding UAVs for multihop underwater communication is still an unexplored area of research.
 \section{Use Cases of UAV-Aided Maritime Communication}
This section covers various use cases of UAV-aided maritime communication such as relaying, wireless power transfer, data offloading, and localization. In the following, we discuss each of the use cases in detail.
\subsection{UAV-based Relaying}
 UAV-based communications are getting growing importance for many applications, particularly with the arrival of high-altitude long-endurance platforms. These UAVs can enable BLoS communications in support of a range of maritime activities. The UAV-based airborne relay will enable range extension for maritime communication services. Also, with the flexible mobility and high possibilities of LoS air-to-sea links, UAVs-enabled relays can display increasingly important advantages for maritime networks as shown in Fig.~\ref{sea2}. 
 \subsection{UAV-Aided Maritime IoT Data Harvesting}
Underwater sensor networks have attained a lot of research intention in recent years, however, it is evident that major obstacles remain to be solved. Several telemetry activities for maritime monitoring, research, and exploitation can be performed based on collecting data from marine buoys rapidly and in real-time. Satellites, ships, and airplanes can all collect marine data, but satellite transmission is often expensive and bandwidth-limited, while manned ships/aircraft have high manpower/mission costs and risks. Therefore, using UAVs that can resist strong winds over the sea surface as an agile data collector appears to be an exciting solution. UAVs can fly near the buoys and use a stable communication channel to wirelessly and quickly capture a significant amount of data because of their high mobility.
\subsection{Maritime Wireless Power Transfer}
Wireless charging has been acknowledged as a viable technology to provide energy supply for battery-limited nodes, such as underwater Internet-of-things (IoT) devices and sensors. UAVs-based wireless charging can bring more flexibility in terms of mobility and accessing hard-to-reach areas \cite{uav8}. Due to the LoS linkages between the UAV and sensors, the UAV-enabled wireless power transfer system may substantially improve energy transfer efficiency by deploying the UAV as a mobile energy transmitter. 
\subsection{Maritime Computation Offloading}
Because of great sensitivity to time and energy consumption, many computation- and data-intensive jobs are challenging to accomplish on maritime energy-constrained devices. UAVs-based mobile edge computing (MEC) appears to be a promising solution to overcome this challenge, providing ubiquitous internet services for emerging maritime applications such as marine environmental monitoring, ocean resource exploration, disaster prevention, and navigation. As a result, UAV-based MEC has emerged as a new paradigm that receives a great deal of attention in both academic and industrial sectors.
Nevertheless, increasing demand for large-scale connection and communications, ultra-low information processing latency, and high dependability in delay-sensitive marine applications pose problems for delivering reliable quality-of-service (QoS) in a resource-constrained maritime network. Therefore, UAV-based MEC, which puts processing and storage resources to the network's edge, has proven to be a promising approach. As such, delay-sensitive operations are performed close to ships by placing edge computing servers in the maritime network, reducing reaction time and relieving traffic congestion. Dynamically varying computing and communication resources and diverse QoS requirements in maritime communication environments make resource allocation a difficult task. UAVs, in this sense, can be utilized for computational task offloading and edge computational purposes.
\subsection{Maritime Localization}
Localization plays a significant role in communication in the TaNTIN environment \cite{akhtar2021tantin}. Maritime localization makes use of a ship's measuring devices to determine the location of other nautical targets. Ocean surveillance satellites can take advantage of space and altitude to cover large areas of the ocean, monitor and submarine operations in real-time, and detect radar signals sent by ships.  Nevertheless, the position precision based on satellites may not be satisfactory, especially in unforeseen situations that require high accuracies, such as ocean rescue and non-cooperative (enemy) ship location. In this case, UAVs can be used to improve the localization accuracy of the targets where the UAVs can be controlled remotely \cite{uav10}. Nevertheless, the self-positioning of UAV platforms and the determination of the location of unknown marine targets by UAVs are itself challenges. 
\section {Research Challenges and Directions}
Although there has been a great interest in UAV-aided maritime communication over the past few years, there are various open research issues that should be targeted. In the following, we explore a few of the promising upcoming research challenges for UAVs-aided maritime communication networks.
\subsection{UAV 3-Dimensions Maritime Trajectory Design}
Exploiting the UAV's high mobility is projected to unlock the full potential of UAV-to-sea communications. Various trajectory optimization models exist in the literature that optimizes air-to-sea communications under different UAV configurations. The problems of trajectory optimization are often non-convex, and variants of the successive convex approximation (SCA) technique are used to solve them sub-optimally.  Nevertheless, these SCA-based approaches depend heavily on trajectory initialization and do not explicitly account for the wind effect. Furthermore, for fixed-wing UAVs that must sustain the forward motion to stay in the air, the computational complexity and resulting trajectory complexity make it costly to collect a high volume of data. Therefore, designing an energy-efficient three dimensions (3D) maritime UAV trajectory is very important. 
\subsection{UAV-to-Sea and UAV-to-UAV Interference Management}
For maritime applications, UAVs largely send data in the downlink. Nevertheless, the capacity of maritime-connected UAVs to establish LoS communication with several sea vessels might lead to severe mutual interference among them and to the ships. To overcome this difficulty, additional advances in the architecture of future UAV-based maritime networks such as enhanced receivers, 3D frequency reuse, and 3D beamforming are needed. For instance, because of their capabilities of detecting and categorizing images, deep learning models can be implemented on each UAV to recognize numerous environment elements, such as the location of UAVs and ships. Such a method will enable each UAV to change its beamwidth tilt angle to minimize the ships' interference. \par
 Another solution to minimize the interference is using online path planning for UAVs that accounts for maritime wireless communication and assist in tackling the interference issues together with new advances in the design of the network, such as 3D frequency reuse. This will allow the UAVs to change their movement based on the rate requirements of both aerial UAV-ship and sea-based sensors, thus increasing the overall network performance. \par
Moreover, in streaming applications, UAV trajectory optimization is also critical. In particular, physical layer technologies such as 3D beamforming can be paired with an interference-aware path planning system to provide more efficient communication links for both sea and aerial users. Therefore, some robust UAV-to-Sea and UAV-to-UAV Interference Management techniques should further be explored to increase overall maritime network performance. 
\begin{figure}[t]
\centerline{\includegraphics[width=0.5\textwidth]{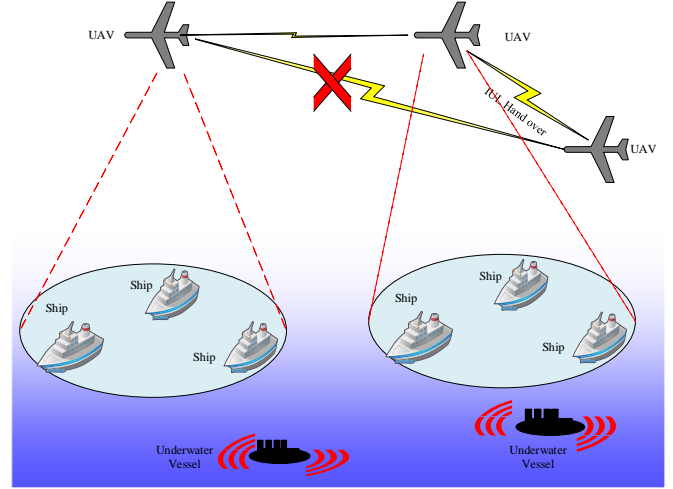}}
 \caption{A depiction of UAV handover in maritime network.}\label{sea3}
\end{figure}
\subsection{3D Mobility Management (3D Handoffs)}
UAVs can be deployed as aerial base stations or aerial users in UAVs-assisted maritime networks. For the case of their deployment as the aerial base stations, 
UAVs can be deployed far away from maritime users such as a ship. This might degrade the signal strength at the receiver ultimately poor mobility performance such as radio connection loss, hand over failure, and even it may cause unnecessary handovers, as shown in Fig.~\ref{sea3}. Additionally, loss of C\&C signal may result in dangerous events such as the collision of UAVs with the commercial aircraft or even it may fall into the sea.\par
For this case, UAVs are deployed as aerial users in the maritime communication networks, they can still face many mobility management issues, especially when there is no LoS link between the maritime base station and the aerial users \cite{uav15}. Although aerial users can still be served by the side lobs of base station antenna, it may cause loss of connection, and handover failure due to lower antenna gains in the side lobs \cite{uav16}. Hence, excellent mobility management for enabling reliable connections between UAVs and ships sailing over the sea level is of essential relevance.
\subsection{Beam-forming for High Mobility Ships and UAVs}
Beamforming and power control design issues are more challenging in maritime communication networks. These design issues are mainly due to frequent switching of frequency access points when the UAVs and/or ships are traveling fast. Moreover, UAVs need to be collaboratively operated because of two smart antenna systems installed on them. This requires two-point power control and two-point beamforming, concurrently. \par
Conjunct power control and beamforming are used to provide reliable coverage by taking the advantage of the location information of the UAVs and ships when the angle of departure (AoD) or angle of arrival (AoA) varies constantly. However, applying a fixed beamforming vector may lead to SINR variations due to variations in AoD and AoA. \par 
Installation of up-tilted UAV antenna and vertical beamforming can improve the coverage for the greater heights. This requires further research, field trials, and simulation to tackle the interference and mobility issues in UAV-assisted maritime networks. More empirical measurements with Doppler effects can be of substantial value for constructing more accurate statistical air–to-sea channel models. Characterizing Doppler effects clearly in channel measurements will be interesting, especially for UAVs and/or ships moving at high speeds.
\section{Conclusion}
This article presents the potentials and challenges of UAV-enabled marine communication networks. The possible architecture of a UAV-based maritime network is identified along with the various types of wireless maritime channel characteristics. Furthermore, several use cases of  UAV-assisted maritime communications are discussed. The article further tries to spur the interests of the researchers on the future evolution of UAVs-enabled maritime communication networks that will enable digital use cases for the future marine economy.
\bibliographystyle{IEEEtran}
\clearpage
\end{document}